\documentstyle[12pt]{article}
\begin{document}
\def\a{\kern+.6ex\lower.42ex\hbox{$\scriptstyle \iota$}\kern-1.20ex a}
\def\be{\begin{equation} }
\def\ee{\end{equation} }
\def\xf{$x_{F}$}
\def\pt{$p_{T}$}
\def\pts{$p_{T, sum}^{2}$}
\def\lc{$\Lambda_{c}^{+}$}
\def\alc{$\overline{\Lambda}_{c}^{-}$}
\def\la{$\Lambda^{0}$}
\def\ala{$\overline{\Lambda}^{0}$}
\def\chikw{$\chi^{2}$}
\def\plmi{$\pm$}
\def\yd{$y_{diff}$}
\def\ydif{$y_{diff}$}
\def\mef{$M_{eff}$}
\def\dfi{$\Delta\phi$}
\def\QCDoathr{$\alpha^{3}_{s}$}
\def\QCDoatwo{$\alpha^{2}_{s}$}
\begin{titlepage}
\begin{flushright}
{\bf BRU/PH/201} \\
arch-ive/9505005 \\
May \ \ 1995
\end{flushright}
\vskip3.5cm
\begin{center}
{\LARGE\bf
Momentum Correlations of Charmed Pairs \\
\vskip0.15cm
Produced in $\pi^{-}-Cu$\ Interactions
\vskip0.15cm
 at 230~GeV/c}
\end{center}
\vspace{0.5cm}
\begin{center}
K.Rybicki$^{a}$ and R.Ry\l ko$^{b}$
\end{center}
\vskip0.2cm
\begin{center}
\footnotesize
\begin{tabular}{ll}
a\ &Institute of Nuclear Physics, PL-30055 Cracow, Poland \\
b\ &Department of Physics, Brunel University, Uxbridge UB8 3PH, UK
\end{tabular}
\end{center}
\vskip1cm
\centerline{ {\bf Abstract} }
\begin{quote}
{\small\rm\hspace*{6mm}  We study the production characteristics of 557 pairs
of charmed hadrons produced in $\pi^{-}-Cu$\ interactions at 230~GeV/c using
a momentum estimator for charmed hadrons with missing decay products.
We find,
the mean value of the transverse momentum squared of the charmed
pairs is  $<$\pts$>=(1.98\pm 0.11\pm 0.09)\;$ GeV$^2$/c$^2$,
the mean rapidity difference is $<|$\yd$|>=0.54\pm 0.02\pm 0.24$,
and the mean effective mass is $<$\mef$>=(4.45\pm 0.03\pm 0.13)\;$ GeV/c$^2$.
Comparing these results with the next-to-leading order QCD predictions we
find an agreement for the \yd\ and \mef, whilst the measured
mean value of \pts\ is significantly larger than the predicted value. }
\end{quote}
\vskip3cm
\centerline{(Submitted for publication in Physics Letters B)}
\end{titlepage}
\setcounter{footnote}{0}
\section{Introduction}
\hspace*{6mm}
Experimental results on charm hadronic production are usually compared
with the  next-to-leading order\footnote{Hereafter called $O(\alpha_{s}^{3})$.}
QCD calculations\cite{NLO_all},\cite{NLO}.
Let us recall that
the ACCMOR\cite{PROD} and the E769\cite{E769} collaborations have shown
that the charm production cross sections, as well as the \xf\ and \pt\
spectra of charmed particles, do not depend strongly on the nature of the
charmed particles\footnote{
Throughout the paper a particle symbol stands for particle and
antiparticle, i.e., any reference to a specific state implies the
charge-conjugate state as well. Thus, e.g.\ $D^{+}\overline{D^{0}}$\
stands for $D^{-}D^{0}$\ as well.}
under consideration ($D^0$, $D^+$, $D^{*+}$, $D^+_s$\ and \lc).
This indicates that the main features of charm production
are determined at the parton level with a relatively small dependence on
the light quarks forming a charmed hadron. On the other hand, the above
characteristics are not very sensitive to the difference between the
leading order\footnote{Hereafter called $O(\alpha^{2}_{S})$.}
and $O(\alpha^{3}_{S})$\ QCD calculations. The charm pair correlations
are more sensitive to this difference.
$O(\alpha^{2}_{S})$\ QCD predicts that the heavy quark pair is produced
exactly in a back-to-back configuration, corresponding to an azimuthal angle
difference of \dfi=$180^{0}$\ and \pts=0. The third order terms cause the
broadening of these distributions. Thus, the distributions of the charmed pairs
test the importance of the next-to-leading order terms. On the other hand,
disagreement with the perturbative results may show the importance of various
nonperturbative effects, which are likely to play an important role
at the energy scale set by the charmed quark mass.

There is a relative abundance of data on the azimuthal angle difference
distribution \dfi\
of a charmed pair in hadroproduction as this angle can be determined from
the directions of the charmed particles without the momentum measurements,
which demand the detection of all decay products. The results of the
WA75\cite{WA75A}, E653\cite{E653B}, ACCMOR\cite{AZI}, and to some extent,
of the WA92\cite{WA92} collaborations show the back-to-back enhancement of
the \dfi\ distribution to be much weaker
than the $O(\alpha_{s}^{3})$ predictions.
There are less experimental results available for distributions of other
kinematical
variables of a charmed pair because they require a knowledge of the
momenta of the
charmed particles, while the fully reconstructed decays
(no missing decay products)
are much less numerous. The E653\cite{E653B,E653C} collaboration
attempted to overcome this difficulty by using a momentum estimator. In this
letter we perform a similar analysis of the ACCMOR data with the aim of
determining the following characteristics of the pair of charmed particles:\\
\pts\ - square of the vector sum of transverse momenta;\\
\ydif\ - the rapidity difference between particles;\\
\mef\ - effective mass of the pair.\\
Since the ACCMOR experiment, data processing and selection of double charm
events have already been described (see ref.\cite{AZI} and references
therein), we recall here only the essential features of the experiment and
the data analysis. Next, we discuss the momentum estimator and its errors.
Finally, the results are presented and compared to the $O(\alpha_{s}^{3})$
QCD predictions.
\section{Experiment, data analysis and acceptance corrections}
\hspace{6mm}
The second phase of the NA32 experiment was performed at
the CERN-SPS using a~negative $230~GeV/c$ beam (96\% pions and 4\%
kaons) and a 2.5mm Cu target. Charm decays were reconstructed with an
improved silicon vertex detector and a~large acceptance spectrometer.
The latter consisted of two magnets, 48 planes of drift chambers and
three multicellular Cherenkov counters allowing $~\pi$, $K$, $p~$\
identification in a wide momentum range. The vertex detector consisted
of a beam telescope (seven microstrip planes) as well as a vertex telescope
with two charge-coupled devices (CCDs) and eight microstrip planes.
The high precision vertex detector allowed the clean reconstruction of
charm decays with very few background events and a purely topological
charm search, which was restricted neither to a limited number of
decay modes nor to any mass window.\\
\hspace*{6mm}Events with the primary vertex inside the target and at least
two secondary tracks not originating from the vertex were selected for
further analysis. These tracks were then used as a seed
to  search for one or more secondary vertices. Events with two secondary
vertices were next selected.
Since most of them involved unseen neutral decay products, one
could not demand the effective mass of charged decay products to be
compatible with the mass of a charmed particle.
Secondary interactions were
excluded by demanding  a separation of the secondary vertex of
at least two standard deviations from the target edge as well as
from the edges of both CCDs. The decays of strange particles were
rejected by demanding that the effective mass $m_{vis}$ of
the charged decay products for $\pi^{+}\pi^{-}$ vertices had to be
larger than the kaon mass. Similarly, all three-prong
vertices, which could be interpreted as an accidental overlap of a
$K^{0}_{S}\rightarrow\pi^{+}\pi^{-}$ (or $\Lambda^{0}\rightarrow p\pi^{-}$)
decay and a~track, were rejected by checking the effective mass of the
$\pi^{+}\pi^{-}$ (or $p\pi^{-}$) combinations. Additionally, the visible
transverse momentum $p^{vis}_{T}$, with respect to the direction of the
parent charmed particle P, was demanded
not to exceed the maximum transverse momentum kinematically allowed
for any particle in the decay channel under consideration. Finally,
decays attributed to short-lived $D^{0}$, $D^{+}_{s}$ or
$\Lambda^{+}_{c}$ had to occur before the second CCD (20 mm from the
target).

In ref.\cite{dbt} such selected secondary vertices were assigned
to various decay modes of $D^{0}$, $D^{+}$, $D^{+}_{s}$
and $\Lambda^{+}_{c}$. This was done with the help of $m_{vis}$
and of the neutral mass defined as
\begin{equation}
m^{2}_{0}=m^{2}_{P}+m^{2}_{vis}-2m_{P}\sqrt{m^{2}_{vis}+(p^{vis}_{T})^{2}}\; .
\end{equation}
For the purpose of this letter the exact assignment is not fully
needed since we only want to identify the charmed particle~P,
as a knowledge of the decay channel is irrelevant here.
To use the momentum estimator we need to know the charmed hadron
mass, its flight vector and the momenta of visible decay products.
As stated in ref.\cite{AZIold} the wrong assignment
can amount to about $5\%$ for $D^{0}$ as well as for $D^{+}$ and to about
$15\%$ for $D^{+}_{s}$. In the same paper it was shown that the observed
number of $D\overline{D}$ events is consistent with that expected from
the charm production cross section, branching fractions, and our
acceptance.\\
\hspace*{6mm}The resulting sample of 557 pairs was used for a study
of azimuthal angle difference \dfi\ and pseudorapidity gap
distributions\cite{AZI}.
Since the production characteristics depend only slightly on the nature
of the charmed hadrons, we ignore this dependence in the present study.\\
\hspace*{6mm}The simulation of the geometrical acceptance of our apparatus
and of the selection criteria requires a complex Monte Carlo program.
This program generates an uncorrelated pair of charmed particles
$P_{i}$ and $P_{j}$.
The particles decay into the observed channels, and we calculate the
acceptance $A_{ij}$ for each combination of decays, each decay being
generated from phase-space distribution. The distributions of the
$x_{F}$ and of the transverse momenta\cite{PROD} of the $P_{i}$\
and $P_{j}$, as well as the lifetime of $\Lambda^{+}_{c}$\cite{LFT} are
taken from this experiment, and the lifetimes of the charmed mesons are taken
from RPP\cite{RPP}. The generated pair of charmed particles is mixed
with tracks from an event randomly chosen from an interaction-trigger
sample recorded with our apparatus, thus faking a real event. All
tracks from such an event are subsequently traced through the apparatus
and subjected to the same cuts as those made in our analysis.
The acceptance $A_{ij}$ vanishes in the backward centre-of-mass hemisphere,
otherwise it amounts to $(1-5)\%$ depending on the decay channel.
The acceptance depends only slightly on \pt, with a small decrease for
increasing \pt. All results presented in this letter have been corrected
for the acceptance.
\section{Momentum estimator}
\hspace{6mm}
The momentum estimator, used in order to account for the unseen decay
products, was first applied to study differential cross sections\cite{E653C}
and the correlations of charmed hadrons\cite{E653B} in the E653 experiment,
where the charm decay vertices were observed in nuclear emulsion.
The NA32 system of the beam hodoscope and the silicon microstrip detectors
supplemented with two CCDs yielded a comparable precision of measuring
the primary and the  decay vertex, thus  allowing us to find
the direction of flight of the charmed hadron with a high level of accuracy.
The visible decay products were identified and their momenta were measured
with a high level of accuracy by the NA32 large-angle spectrometer.
The estimation of the
charmed hadron momentum is based on the assumption that in its rest frame
the invisible system momentum is perpendicular to the
charmed hadron laboratory momentum.
This assumption, the angle between the
charmed hadron flight vector, and the visible momentum vector,
uniquely  fix the boost from the charmed hadron rest frame to the laboratory
frame, thus allowing calculation of the laboratory momentum of the charmed
hadron. The momentum estimator was obviously not used for fully reconstructed
decays in our data sample.\\
\hspace*{6mm}The systematic error of the momentum estimator has been determined
using a special procedure\cite{decest}. The charmed hadrons have been generated
according to the distributions measured in the NA32 experiment\cite{PROD}.
Then, the hadron undergoes the multibody isotropic decay with momenta chosen
according to the phase-space distribution\footnote
{We belive this is a good approximation in our case since, the majority of
our charmed
hadrons are pseudoscalar mesons, and the main source of  error of the momentum
estimator arises from the angular distribution of the decay products, rather
than a particular choice of their momenta. }.
Next, the momentum of the visible decay products and the charmed hadron
flight vector are found, and the momentum estimator is applied.
Then, various kinematical variables for the charmed pair are calculated,
using the estimated momentum, and finally they are compared to the
generated variables. The errors for various decay modes are collected in
Table 1. The systematic errors of various kinematical variables
are determined for seven decay modes, which constitute a fair majority of
events in our sample.  The distributions of the estimated minus the  generated
kinematical variables of the charmed pairs are symmetric and centered
at zero. An example of such distributions of the differences are plotted in
Fig.\ 1 for the
$D^0\rightarrow K^-\pi^+(\pi^0\pi^0)$\ mode with unseen $\pi^0$'s.
This decay channel is almost twice as abundant in our double charm sample
than any other decay mode listed in Table 1. In the second column of Table 1
we also quote the estimation error of the laboratory momentum of the
charmed meson, relative to the generated one. As expected, the
error increases with the increasing fraction of energy carried by invisible
decay products. Calculating the weighted average we obtain 15.6\%\ accuracy
for our estimate of the laboratory momentum. It should be noted that this error
has little impact on the reconstructed kinematical variables of the charmed
pairs, especially on the \pts\, for which there is very little dependence
on the decay channel. We use the RMS of the distribution of the difference
between the estimated and
the generated variable as the measure of the systematic error on each
kinematical variable in Table 1, checking that the mean value of the
difference distribution is negligible in each case.
The overall errors are calculated using the weights based on the numbers
of such double charmed events observed in the ACCMOR experiment\cite{dbt}.
We assume these errors to be valid also for the remaining, less abundant,
decay modes.
\begin{table}
\begin{center}
Table 1 \\
\vspace*{0.7ex}
\begin{tabular}{||l||c||c|c|c||} \hline\hline
$D^+/D^0$\ decay & $\Delta p_{lab}/p_{lab}$ & $\Delta$\pts\  & $\Delta$\yd\
& $\Delta$\mef\   \\
mode  &  $[$ \%\ $]$    & $[$GeV$^2$/c$^2 ]$  & $[$ 1 $]$
& $[$ GeV/c$^2 ]$   \\
\hline\hline
$K^-\pi^+ (\pi^0)$           & 15.1  & 0.086  & 0.229 & 0.124 \\
$K^-\pi^+\pi^- (\pi^0)$      & 10.0  & 0.078  & 0.157 & 0.099 \\
$(K^0)\pi^+\pi^-$            & 18.7  & 0.094  & 0.276 & 0.143 \\
$(K^0)\pi^+\pi^+\pi^-$       & 14.2  & 0.087  & 0.222 & 0.123 \\
$K^-\pi^+ (\pi^0\pi^0)$      & 15.8  & 0.089  & 0.244 & 0.137 \\
$(K^0\pi^0)\pi^+\pi^-$       & 19.6  & 0.095  & 0.292 & 0.148 \\
$(K^0\pi^0)\pi^+\pi^+\pi^-$  & 14.9  & 0.090  & 0.230 & 0.129 \\
\hline\hline
Average                      & 15.6 & 0.089 & 0.238 & 0.130 \\
\hline\hline
\end{tabular}
\end{center}
Table 1. The systematic errors of the momentum estimator for various decay
modes (with unseen decay products in brackets). The numbers are RMS values of
the distributions of the estimated minus the generated kinematical variable
of the charmed
pairs. The average errors are calculated using weights proportional to the
abundance of the decay channel in our double tag sample.
\end{table}

\section{Results}
\begin{table}
\begin{center}
Table 2 \\
\vspace*{0.7ex}
\begin{tabular}{||c||c|c|c|c|c||} \hline\hline
Experiment  & $\sqrt{s}$ & No of & $b_p$
                                   & $\sigma_y$ & $\alpha_M$ \\
Beam  & $[$GeV$]$  & pairs & $[($GeV/c$)^{-2}]$ & $[1]$
                                    & $[($GeV/c$^2)^{-1}]$ \\
\hline \hline
&&&&&\\
E653\ $p-emul.$ & 38.7 & 35 & --- & $1.85^{+0.45}_{-0.41}$ &
                                               $0.53^{+0.14}_{-0.10}$ \\
800 GeV/c \cite{E653B} &&&&&\\
&&&&&\\
\cline{1-6} \cline{1-6}
&&&&&\\
WA75\ $\pi^{-}-emul.$ & 25.6 & 177 & 0.50\plmi 0.10 & 1.00\plmi 0.06$^{a)}$ &
                                               $1.17^{+0.13}_{-0.17}$ \\
350 GeV/c \cite{WA75B} &&&&&\\
&&&&&\\
\cline{1-6} \cline{1-6}
&&&&&\\
NA32\ $\pi^{-}-Cu$ & 20.8 & 557 & 0.50\plmi 0.02 & 0.65\plmi 0.02 &
                                               1.39\plmi 0.06   \\
230 GeV/c &  &  &  & & \\
\chikw /NDF & & & 26.1/18 & 12.5/9 & 8.1/5 \\
&&&&&\\
\hline\hline
\end{tabular}
{\small $^{a)}$~calculated from the measured mean value of the $|$\ydif$|$\
distribution.}
\end{center}
Table 2. The results of the maximum-likelihood fits to the NA32 data
and the corresponding results of other experiments.
\end{table}

\hspace*{6mm}Using the laboratory momenta estimated in the previous section
we calculate the characteristics of charmed pairs in our sample.
The acceptance corrected \pts\ distribution is shown in Fig.\ 2.
We have fitted the \pts\ distribution with
\be
\frac{d\sigma}{dp_{T, sum}^{2}} \sim e^{-b_p\; p_{T, sum}^{2}} \;
\ee
using the maximum-likelihood method. We have done a similar fit for the
distribution of the rapidity difference of the charmed pairs
\be
\frac{d\sigma}{dy_{diff}} \sim e^{-\frac{y_{diff}^2}{2\sigma_y^2}} \; .
\ee
In Fig.\ 3 we plot the absolute value of the rapidity
difference $|$\ydif$|$. Finally, the distribution of the effective mass
\mef\ of charmed particles is fitted to
\be
\frac{d\sigma}{dM_{eff}} \sim e^{-\alpha_M\; (M_{eff}-M_0)} \;  ,
\ee
where $M_0$=$2\, m_{D}$.
The \mef\ distribution is shown in Fig.\ 4. The fitted parameters, their
statistical errors, the \chikw\ values, and the NDF (number of degrees of
freedom) are collected in Table~2. The results are compared with the results
of other charm hadroproduction experiments.
As can be seen, the \ydif\ and \mef\ distributions become broader with
increasing $\sqrt{s}$, in accordance with QCD predictions.
\section{Comparison with $O(\alpha_{s}^{3})$ QCD predictions}
\hspace*{6mm}
The next-to-leading order QCD predictions for the charmed pairs production,
calculated using the program from ref.\cite{NLO}, are collected in Table 3.
The mean values of the distributions\footnote
{The \QCDoathr\ predictions displayed in Table 3 are the mean values
calculated using the program from ref.\cite{NLO}, and  its (signed)
weights. These mean values are consistent with the mean values
determined from the histograms produced by the program, with
the maximum number of bins allowed.}
were calculated using HMRSB\cite{HMRSB} parametrization
of the nucleon and the central set SMRS\cite{SMRS} parametrization of the
pion structure functions\footnote
{It should be noted that the results are
not very sensitive to the structure functions, e.g.,
taking other (SMRS) pion structure functions results in
\plmi 0.01 GeV$^2$/c$^2$, \plmi 0.02,
and \plmi 0.02 GeV/c$^2$\
differences for the mean values of the \pts, \yd\ and \mef\ distributions of
the charmed hadron pairs, respectively.}.
The default value of $\Lambda_{\overline{MS}}^5$=122 MeV was used, and
the $\mu_R$\ and $\mu_F$\ scales were chosen to be $m_c$\ and 2$m_c$,
respectively, where $m_c$=1.5 GeV/c$^2$.
The first row in Table 3 represents the results for bare charmed quarks
with \xf$>$0.
As the fragmentation generally leads to the softening of the distributions,
we use the hard fragmentation function
\be
f(z)\;  \sim \; \delta (z-1)\; ,
\ee
where $z=E_D / E_c$, is the fraction of charmed quark energy transferred to the
charmed hadron. The results ($x_{F}>0$\ for both charmed hadrons)
are presented in the second row of Table 3.
Our results, this time determined as the acceptance weighted mean values
of each distribution, are shown in the last row of Table 3.
First errors are the statistical errors of the mean values.
The systematic errors are taken from the bottom row of Table 1.\\
\begin{table}
\begin{center}
Table 3 \\
\vspace*{0.7ex}
\begin{tabular}{||c||c|c|c||} \hline\hline

Result & $<$\pts$>$        &  $<|$\yd$|>$   & $<$\mef$>$  \\
       &  $[$GeV$^2$/c$^2 ]$ & $[$ 1 $]$      & $[$GeV/c$^2 ]$ \\
\hline\hline
&&&\\
\QCDoathr\ & 0.434 & 0.631 & 3.851 \\
&&&\\
\QCDoathr\ + h.f.  &  0.396 & 0.559 & 4.467 \\
 &&&\\
\hline\hline
&&&\\
NA32   & 1.98\plmi 0.11\plmi 0.09 & 0.54\plmi 0.02\plmi 0.24
& 4.45\plmi 0.03\plmi 0.13 \\
&&&\\
\hline\hline
\end{tabular}
\end{center}
Table 3. The $O(\alpha_{s}^{3})$ QCD predictions (first row) supplemented by
the hard charmed quark fragmentation (second row) compared to the NA32
experimental results (bottom row). The first error
is the statistical error of the mean
value of the distribution. The second error is the systematic error due to
the uncertainties of the momentum estimator.
\end{table}
\hspace*{6mm}For the \mef\ and \ydif\ distributions there is  a good agreement
with the $O(\alpha_{s}^{3})$ QCD supplemented  by the hard hadronization of
the charmed quark. On the other hand, the experimental value of \pts\ is much
larger than the predicted value. Let us recall that this result depends
weakly on the uncertainties associated with the momentum estimator (cf.\
Table 1). It should also be mentioned that $<$\pts$>$ = (1.52\plmi0.34)
GeV$^2$/c$^2$ for 20 pairs of fully reconstructed charmed particles in this
sample\cite{AZI}. Moreover, the WA75 collaboration\cite{WA75B} measured
$<$\pts$>$ =~(2.00$^{+0.50}_{-0.33}$)\ GeV$^2$/c$^2$. All these results are
qualitatively consistent with the relatively flat \dfi~distributions
described previously.\\
\hspace*{6mm} To conclude, the perturbative next-to-leading order QCD
calculation, whilst quite successful in determining total cross sections,
single particle spectra and "longitudinal" correlation distributions
of charmed pairs (\ydif\ or \mef),
is not able to reproduce the "transverse" correlation distributions
(\pts~and \dfi).
As discussed in refs.\cite{AZI,Rid} a fairly large
intrinsic (or "primordial") transverse momentum of the colliding quarks,
comparable with the measured mean value of \pts, is needed to reproduce
the experimental results.
\begin{flushleft}
{\large\bf Acknowledgements}
\vskip0.1cm
\end{flushleft}
\hspace*{6mm} The authors are much indebted to the ACCMOR collaboration for
the kind permission to use their data.

\newpage
\vskip2cm
{\large\bf Figure captions: }\\
\vskip0.5cm

Fig.1.  Plots of the estimated minus the generated kinematic
variables of charmed pairs like: the \pts\ {\bf (b)}; the \ydif\ {\bf (c)};
and the \mef\ {\bf (d)}; with one decay vertex being
$D^0\rightarrow K^-\pi^+(\pi^0\pi^0)$\ mode ($\pi^0 \pi^0$\ system unseen).
The same difference is plotted for the laboratory momentum {\bf (a)}
of a single $D^0$\ decaying as above.\\

Fig.2. The acceptance corrected distribution of the vector
sum of transverse momenta squared of the charmed hadron pairs \pts.
The points are experimental data, the solid line is the result of the
maximum-likelihood fit, with $b_p$=(0.50\plmi 0.02) (GeV/c)$^{-2}$.\\

Fig.3. The acceptance corrected distribution of the charmed
hadron pairs
rapidity difference $|$\ydif$|$. The points are experimental data, the solid
line is the result of the maximum-likelihood fit, with
$\sigma_y$=0.65\plmi 0.02.\\

Fig.4. The acceptance corrected distribution of the charmed
hadron pairs invariant mass \mef. The points are experimental data, the solid
line is the result of the maximum-likelihood fit, with
$\alpha_M$=(1.39\plmi 0.06) (GeV/c$^2)^{-1}$.
\end{document}